\documentclass[aps,prl]{revtex4-2}

\usepackage{graphicx}
\usepackage{dcolumn}
\usepackage{lineno}
\usepackage{ulem}
\usepackage{bm}

\usepackage{xcolor}
\usepackage{soul}

\usepackage{physics}
\usepackage{comment}

\begin{document}

\preprint{APS/123-QED}

\title{Generalized Lotka-Volterra Systems with Time Correlated Stochastic Interactions }

\author{Samir Suweis$^{1,2,3}$}
\author{Francesco Ferraro$^{1,2,4}$}
\author{Christian Grilletta$^{1}$}
\author{Sandro Azaele$^{1,2,4,*}$}
\author{Amos Maritan$^{1,2,4,*}$}

\affiliation{$^1$Laboratory of Interdisciplinary Physics, Department of Physics and Astronomy ``G. Galilei'', University of Padova, Padova, Italy}
\affiliation{$^2$ INFN, Sezione di Padova, via Marzolo 8, Padova, Italy - 35131}
\affiliation{$^3$ Padova Neuroscience Center, University of Padova, Padova, Italy}
\affiliation{$^4$National Biodiversity Future Center, Piazza Marina 61, 90133 Palermo, Italy}
\affiliation{$^*$These authors contributed equally to the work}

\date{\today}

\begin{abstract}
In this work, we explore the dynamics of species abundances within ecological communities using the Generalized Lotka-Volterra (GLV) model. At variance with previous approaches, we present an analysis of stochastic GLV dynamics with temporal fluctuations in interaction strengths between species. We develop a dynamical mean field theory (DMFT) tailored for scenarios with annealed colored noise and simple functional responses. We show that time-dependent interactions can be effectively modeled as environmental noise in the DMFT and we obtain analytical predictions for the species abundance distribution that well matches empirical observations. Our results suggest that environmental noise favors species coexistence and allows to overcome the complexity-stability paradox, especially in comparison to dynamics with quenched disorder. This study offers new insights not only into the modeling of large ecosystem dynamics, but also proposes novel methodologies for examining ecological systems.

\end{abstract}

\maketitle

\section{\label{sec:level1}Introduction}
In the last few years, the approach of statistical physics has decisively contributed to our understanding of ecological processes by providing powerful theoretical tools and innovative steps towards the comprehension and synthesis of broad empirical evidence \cite{azaele2016statistical}, especially in microbial ecology \cite{quemener2014thermodynamic,xiao2017mapping,grilli2020macroecological,hu2022emergent}. In fact, the advent of next-generation sequencing techniques is generating an increasing volume of ecologically relevant data, characterizing several microbial communities in different environments and involving a large number of species \cite{integrative2019integrative,gilbert2019community}. For this reason, the theoretical approach has shifted from traditional dynamical systems theory, which is effectively used for a small number of species \cite{reichenbach2007mobility,friedman2017community}, to statistical physics based methods that are better suited for large systems \cite{tikhonov2017collective,tu2017collapse,marsland2020minimal,peruzzo2020spatial,batista2021path,gupta2021effective}. Given the inherent unknowns in species interactions, several recent works have proposed modeling species interactions through Generalized Lotka-Volterra equations with quenched random disorder (QGLV), where the underlying interaction network is fully connected, leading to a number of interesting results \cite{barbier2018generic,biroli2018marginally,galla2018dynamically,pearce2020stabilization}. The phase diagram of these models describes a system with transitions from a single stationary state to a non-physical unbounded one, passing through a multiple-attractors state as the variance of the interactions between species exceeds a critical value \cite{bunin2017ecological,galla2018dynamically, biroli2018marginally}. Furthermore, the addition of demographic noise to the QGLV leads to new phases such as the Gardner phase \cite{altieri2021properties}.

The QGLV model assumes that species interactions remain constant over time. However, empirical ecological systems are characterized by temporal fluctuations in species interactions, influenced by variations in environmental conditions, resource availability, and other factors that operate on a timescale comparable to population dynamics \cite{thompson1999evolution,suweis2013emergence,fiegna2015evolution,ushio2018fluctuating,pacciani2021constrained}. 
In the global equilibrium phase, the biodiversity of the QGLV model is limited by the stability-diversity paradox \cite{mccann2000diversity,allesina2015stability,gibbs2018effect,biroli2018marginally}. Moreover, the species abundance distribution (SAD), as obtained in the limit of a large number of species within the dynamical mean field theory (DMFT) or the cavity method, is a truncated Gaussian \cite{bunin2017ecological,galla2018dynamically}, very different from the heavy tail SAD observed in empirical microbial \cite{sala2016stochastic,ser2018ubiquitous,grilli2020macroecological} or in forest\cite{azaele2006dynamical} communities. 

In the present study, within the established framework of the GLV model featuring a fully connected random interaction network, we explicitly consider time-dependent species interactions and a Monod functional response, commonly used for modelling the growth of microorganisms \cite{monod1949growth}.

Specifically, we adopt the hypothesis that these interactions can be modeled as stochastic colored noises, which we call annealed GLV (AGLV).
Strikingly, the introduction of temporal stochastic fluctuations in the strengths of species interactions yields results that are remarkably rich and ecologically relevant, filling the above mentioned gaps for the QGLV. 


We emphasize that, while the use of GLV equations within a fully connected interaction network for the modeling of complex ecosystems may appear unrealistic, the emergent characteristics are expected to persist even in more realistic scenarios where analytical treatments are inherently intricate and often unattainable.
Moreover, theoretical models as the GLV, stand as the scaffolding upon which empirical research is built, offering a controlled setting where foundational ecological mechanisms can be disentangled. As such, they serve as a crucible for testing the robustness and generalizability of ecological theories.
Here, in particular, we want to understand the effect of including temporal variations of the species interactions strengths. The insights gained can direct and refine further empirical inquiry, setting the stage for a productive interplay between theory and empirical observations.

Let us consider $x_i(t)$, the population at time $t$ of the species $i$. Then the dynamics of the AGLV system with colored noise for $S$ interacting species are given by
\begin{equation}
\dot{x}_i(t)=x_i(t)\big[r_i\left(1-x_i(t)/K_i\right)+\sum_{j\neq i}\alpha_{ij}(t)J(x_j(t))+h_i(t)\big],
\label{eq:glv}
\end{equation}
with $i=1,...,S$, and where  $\alpha_{ij}(t)=\mu/S+\sigma z_{ij}(t)/\sqrt{S}$ for $i\neq j$ and $\{z_{ij}(t):t>0\}$ are independent Gaussian random variables  with $\overline{z_{ij}(t)}=0$, $\overline{z_{ij}(t)z_{kl}(t')}=\delta_{i,k}\delta_{j,l}P(\Delta t|\tau)=\frac{1+2\tau/\tau_0}{2\tau}e^{-\Delta t/\tau}$, where $\delta_{x,y}$ is the Kronecker delta and $\Delta t=|t-t'|$;  $J(x)$ is a generic function of $x$ and in particular we will consider two cases: $J(x)=x$ -- to make direct comparison with QGLV -- and $J(x)=\frac{x}{1+c x}$, with $c$ known as handling time \cite{papanikolaou2021predators}. $h_i(t)$ is a possible time-dependent external field.  For simplicity, we set $\tau_0, r_i, K_i=1$ for $i=1,...,S$ and work with dimensionless variables/parameters.
From this general annealed formulation with colored noise, the limit $\tau\rightarrow 0$ corresponds to the white noise (AWN) dynamics. 
In Fig. 1 we show the effect of time-correlated noise on the species abundance evolution for $J(x_j(t))=x_j(t)$.  In our model fluctuations scale like the size of the population, generating a time-dependent multiplicative noise that keeps the trajectories away from zero, thus preventing their extinction for any finite $\tau$. This behavior is substantially different from the one of models with quenched noise, where all species’ populations with a negative growth rate are doomed to extinction in the infinite time limit.
Species populations undergo recurring quasi-cycles of high and low abundances, whose average frequency depends on the value of $\tau$. This cyclic behavior is instrumental in promoting the coexistence of multiple species within the ecosystem and is present for all ranges of finite $\tau$, including the limit $\tau\rightarrow 0$. As we show, the same results hold for $J(x_j(t))=\frac{x_j(t)}{1+c x_j(t)}$.
Facilitating species coexistence through cyclic fluctuations is a mechanism that has also been observed in the chaotic phase of the QGLV \cite{pearce2020stabilization,roy2020complex,dePirey2023}.
\begin{figure*}[h]
\includegraphics[width=\textwidth]{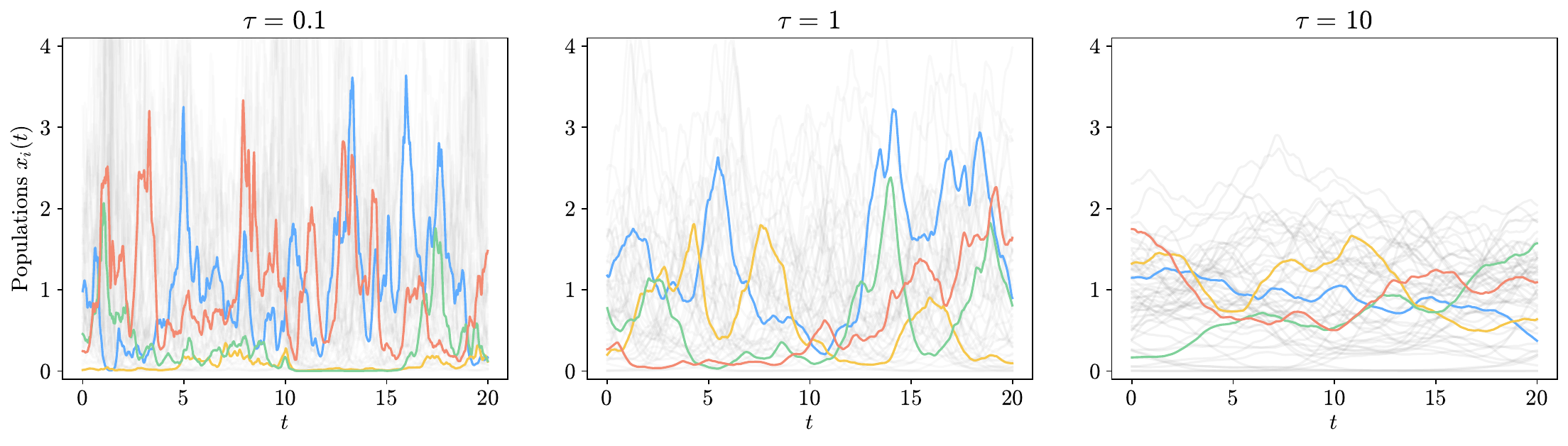}
\caption{\label{fig1}Examples of species abundances trajectories obtained by simulating eq.~(\ref{eq:glv}) for $S=100$, $h=0$ and different values of the characteristic correlation time $\tau$. A) $\tau=0.1$, $\mu=0$, $\sigma=1$; B) $\tau=1$, $\mu=0$, $\sigma=0.8$; C) $\tau=10$, $\mu=0$, $\sigma=0.6$. Cyclic behavior is instrumental in promoting the coexistence of multiple species within the ecosystem and is present for all ranges of $\tau$, including the limit $\tau\rightarrow 0$. Different values of $\sigma$ as $\tau$ varies are chosen to better visualize population fluctuations as a function of time. Additional figures are given in the Supplementary Methods for other values of the parameters $\mu,\sigma$ and $\tau$.}
\end{figure*}

The DMFT for the general AGLV eq.~\eqref{eq:glv} is given by (see Supplementary Methods)
\begin{equation}\label{eq:dmft}
\dot{x}(t)=x(t)\big[1-x(t)+\mu M(t)+\sigma \eta(t)+h(t)\big],
\end{equation}
where $M(t)=\mathrm{E}(x(t))$ and in the following we set $h(t)=0$. The self-consistent Gaussian noise $\eta(t)$ is such that $\mathrm{E}(\eta(t))=0$ and $\mathrm{E}(\eta(t)\eta(t'))=P(\Delta t|\tau)\mathrm{E}(J(x(t))J(x(t')))$.

\begin{figure}[t]
\includegraphics[width=0.45\columnwidth]{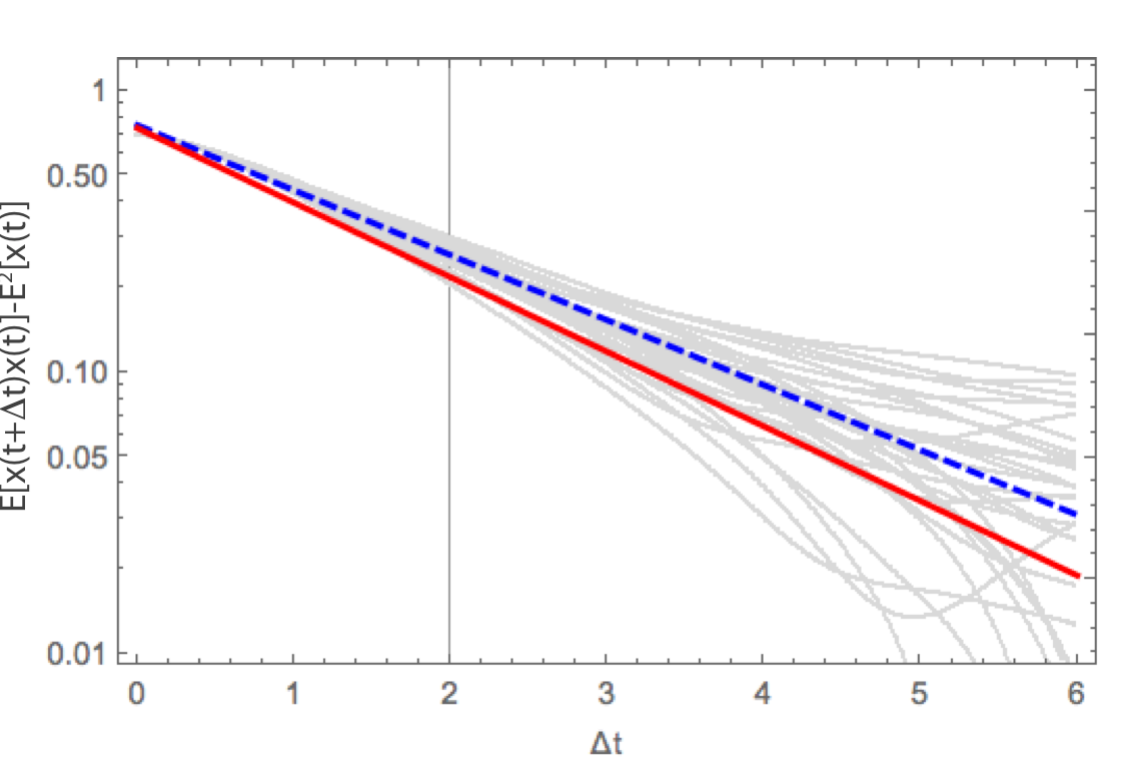}
\caption{\label{fig2} Gray lines represent the empirical auto-correlation function for $J(x)=x$ calculated as $\mathrm{E}(x(t)x(t+\Delta t))-\mathrm{E}^2(x(t))$ (averaged over $t$) for each of the species simulated using the AGLV (here with  $S=30$, $\tau=1$, $\mu=0$ and $\sigma=0.8$). The blue dashed line represents the exponential fit, while the thick red line represents the exponential decay given by the ansatz eq.~\eqref{eq:ansatz_corr} using the parameters inferred from $P_{\tau}^*(x)$ (see also Fig. \ref{fig3}B). For this set of parameters and for $J(x)=x$, we find $\tau_x=2$, and thus we expect that our ansatz works for $\Delta t<2$ (i.e., on the left of the vertical line indicating $\Delta t=2$). A similar procedure has been applied for $J(x)=x/(1+cx)$ (see Supplementary Figures).}
\end{figure}

From Fig. \ref{fig2} we can see that at stationarity, the (connected) auto-correlation function of $J(x(t))$ has an exponential-like decay
\begin{equation}
    \mathrm{E}(J(x(t))J(x(t')))-\mathrm{E}^2(J(x))\approx \big(\mathrm{E}({J(x)}^2)-\mathrm{E}^2(J(x))\big) e^{-\frac{|\Delta t|}{\tau_x}},
    \label{eq:ansatz_corr}
\end{equation}
 and exploiting eq.~(\ref{eq:ansatz_corr}) we can simplify the self consistency for $\eta$ as  $\mathrm{E}(\eta(t)\eta(t'))=P(\Delta t|\bar{\tau})\mathrm{E}({J(x)}^2)$, at least in the relevant regime $|\Delta t|=|t-t'|\ll \tau_x$, with the new effective time scale $\bar{\tau}=1/\big[1/\tau+\big(1- \mathrm{E}^2(J(x))/\mathrm{E}({J(x)}^2)\big)/\tau_x\big]$ (see Supplementary Methods for further details). 

 With this simplification we can now use the Unified Colored Noise Approximation (UCNA) \cite{jung1987} on eq.~(\ref{eq:dmft}), which, for both cases of $J(x)$ considered here, leads to the same stationary SAD 
\begin{equation}\label{eq:PstazUCNA}
    P_{\tau}^*(x)=\frac{x^{-1+\delta_{\tau}}}{Z}\left(\frac{1}{\bar{\tau}}+x\right)e^{-\frac{x}{D}-\frac{\bar{\tau}}{2D}(x-\bar{x})^2},
\end{equation}
where $x>0$, $Z$ is the normalization constant, that can be computed analytically, 
$\delta_{\tau}=(1+\mu M^*)/D(\tau)$, $D(\tau)=\sigma^2 \mathrm{E^*}(J(x)^2)(1+2\tau)\bar{\tau}(\tau)/2\tau$ and $\bar{x}=1+\mu M^*$ ($\mathrm{E^*}(\cdot)$ denotes the average with the distribution $P_{\tau}^*$ and $M^*=\mathrm{E^*}(J(x))$). 

\begin{figure*}[h]
 \centering 
\includegraphics[width=0.8\paperwidth]{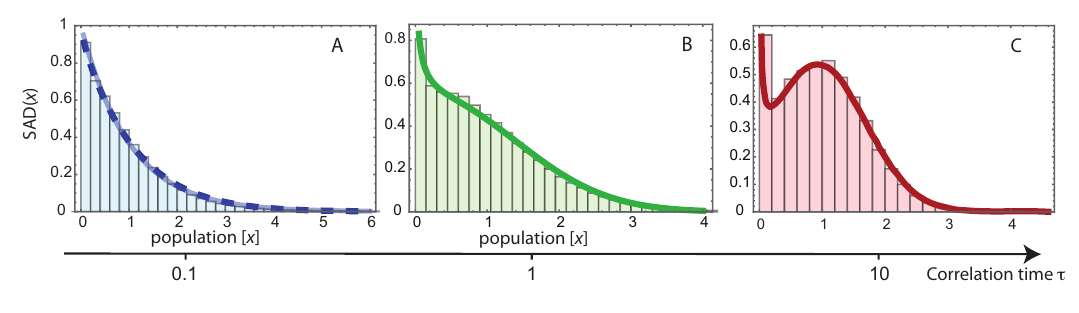}
\caption{Comparison between numerical and analytical solutions. The histograms represent the Species Abundance Distributions (SADs) obtained by simulating the full AGLV system given by eq.~\eqref{eq:glv} for $S=30$ species, the solid lines are the corresponding SADs given by the DMFT and UCNA. A) Eq.~(\ref{eq:PstazUCNA}) with $\tau=0.1$. As blue dashed line we also show the AWN solution given by eq.~\eqref{eq:PstazAnnS}; B-C) Colored noise AGLV (eq.~\eqref{eq:PstazUCNA}) with $\tau=1$ and $\tau=10$, respectively. The analytical DMFT perfectly describes the SAD given by the numerical simulations of the full system even if the number of species is not very high ($S=30$). The initial conditions of the populations in all cases are drawn from $x_i\sim U[0.1,0,2]$, where $U$ denotes the uniform distribution, while the values $\mu$ and $\sigma$ are as in Fig. \ref{fig1}. The different values of $\sigma$ as $\tau$ varies were chosen to better visualize the change of the SAD toward that of a truncated Gaussian plus a Dirac's delta at $x=0$ obtainable in the $\tau=+\infty$ limit.}\label{fig3}
\end{figure*}

UCNA is recognized for its exactness at both $\tau=0$ and $\tau=+\infty$, proving to be a reliable interpolation method for intermediate values of $\tau$ \cite{jung1987}. This is explicitly illustrated in Figure \ref{fig3}, where the analytical solution of the SADs for various $\tau$ values, obtained through both the DMFT and UCNA, is compared with numerical solutions derived from integrating eq.~(\ref{eq:glv}) for a system comprising $S=30$ species with $J(x)=x$ (the same result also holds for the case with the functional response, Supplementary Methods). In the Supplementary Methods, we also perform a sensitivity analysis showing that the analytical solutions match very well the numerical ones for a wide range of parameters.

As demonstrated in both Figures \ref{fig1} and \ref{fig3}, the introduction of time-dependent fluctuations in interactions promotes species coexistence. This is attributed to the induced cyclic behavior, causing the species' growth rate to transition from negative to positive, preventing extinctions, as also verified numerically (see Fig. \ref{fig3}). Only when $\tau=+\infty$ a Dirac's delta develops at $x=0$, whose amplitude represents the fraction of extinct species. Nevertheless, in practical applications, distinguishing rare species from extinct ones in samples with a finite number of species, $S$, requires consideration. This can be addressed by introducing a threshold value, $x_{th}$, such that $P_{\tau}^*(x\leq x_{th})= \int_0^{x_{th}}P_{\tau}^*(x)dx \approx 1/S$.
Notice that $P_{\tau}^*(x)$ is basically an interpolation between a truncated Gaussian (peaked at $x>0$) and a Gamma distribution. The former is known to be the solution for the SAD of the DMFT in the case of random quenched interactions in the single equilibrium phase \cite{bunin2017ecological,galla2018dynamically}, while the latter we will show is the exact solution of the AWN case, corresponding to the limit $\bar{\tau}\sim \tau\rightarrow 0$ of eq.~\eqref{eq:dmft}.
In the AWN limit, in fact, the DMFT equation is the same of eq.~(\ref{eq:dmft}),
but in this case with $\mathrm{E}(\eta(t)\eta(t'))=\Sigma^2(t)\delta(t-t')$, $\Sigma^2(t)=\mathrm{E}(J(x(t))^2)$, and the multiplicative noise term $x(t)\eta(t)$ should be interpreted in the Stratonovich sense \cite{kupferman2004ito}. At stationarity, the self-consistency imposes $M^*=\mathrm{E^*}(J(x))$ and ${\Sigma^{*}}^2=\mathrm{E^*}(J(x)^2)$. The exact stationary distribution $P_0^*$ can be derived from the Fokker-Planck Equation corresponding to eq.~\eqref{eq:dmft} and it reads (for any $J(x)$):
\begin{equation}\label{eq:PstazAnnS}
    P_0^*(x)=\frac{\beta^{\delta}}{\Gamma(\delta)}x^{-1+\delta}e^{-\beta x},
\end{equation}
and it coincides with the limits of $P_{\tau}^*(x)$ when $\tau \to 0$.  We also have $\lim_{\tau \rightarrow 0} \delta_{\tau}=2(1+\mu M^*)/(\sigma^2{\Sigma^{*}}^2)\equiv\delta$ with $M^*=1/(1-\mu)$ and ${\Sigma^{*}}^2=\delta(\delta+1)/\beta^2$ for $J(x)=x$; while for $J(x)=x/(1+cx)$ we find $M^*= \delta\left(-c+e^{\beta/c}\left(c+\beta+c \delta\right) E_{1+\delta}(\frac{\beta}{c})\right)/c^3$ and $\Sigma^*=e^{\beta /c } \delta E_{1+\delta}(\frac{\beta}{c})/c$, where $E_{n}(x)=\int_1^{\infty}e^{-x t}/t^n \mathrm{d}t$ is the exponential integral function.

Therefore, for $J(x)=x$ we can write explicitly the SAD's parameters as a function of $\mu$ and $\sigma$ as (see Supplementary Methods)
\begin{equation}\label{eq:strat_beta}
\beta=\frac{\sigma^2}{2}\delta(\delta+1);\quad
\delta=\frac{2}{\sigma^2}(1-\mu)-1,
\end{equation}
while for the case with functional response, we can simply solve numerically the equations for $M^*$ and $\Sigma^*$ as a function of $\beta$ and $\delta$.

The predicted SADs by eq.~\eqref{eq:PstazUCNA} through the DMFT and UCNA are plotted as continuous lines in Figure \ref{fig3}. In panel (A) we also plot, as a dark blue dashed line, the AWN solution $P^*_0(x)$ given by eq.~\eqref{eq:PstazAnnS}. In this case, the distribution parameters are directly calculated from eq.~\eqref{eq:strat_beta} as a function of $\mu$ and $\sigma$. Instead, for eq.~\eqref{eq:PstazUCNA} and for the solution with $J(x)=x/(1+cx)$, the parameters are obtained by first fitting the distribution and then checking the agreement with the self-consistent equations, and vice versa (error below $5\%$, see Supplementary Methods).

Using the chosen value of the correlation time, $\tau$, the parameters $D(\tau)$ and $\bar{\tau}(\tau)$ given by our analytical framework, we can deduce the value of $\tau_x$  in eq.~\eqref{eq:ansatz_corr}. The red line in Figure \ref{fig2} shows that indeed the predicted value of $\tau_x$ is consistent with the decay of $\mathrm{E}(x(t)x(t'))$ obtained by simulating the full AGLV system given by eq.~\eqref{eq:glv}  (error below $5\%$, see Supplementary Methods also for AGLV with functional response).

To make a comparison between GLV with quenched and annealed interactions, we also investigate the phase diagram for the case $\tau=0$ and $J(x)=x$. Since $\delta>0$ (see eq.~\eqref{eq:PstazAnnS}) and $\mathrm{E}(x)>0$, in order for the stationary solution to exist, we have the conditions $\sigma<\sqrt{2(1-\mu)}$ and $\mu\leq1$, leading to a lower bound for the unbounded growth phase of the AGLV as shown in Fig. \ref{fig4}. However, by solving numerically the self-consistent eq.~\eqref{eq:dmft} (see Supplementary Methods) and also performing the numerical simulation of the entire GLV systems, we find that below this bound, even though a stationary solution exists, it may not be reached. In particular, in the red-purple region of Fig. \ref{fig4}, independently of the initial condition for $x(t=0)$, there is a singularity at finite times, leading to the explosion of the species population. In the light-blue region instead, if we start close to the predicted stationary solution $P^*(x)$, then we always find that the stationary solution is reached and it coincides with the one predicted by the DMFT eq.~\eqref{eq:PstazAnnS}. However, there is a set of initial conditions (for sufficiently large $x(t=0)$) for which $x(t)$ may diverge for finite $t$. Such divergent trajectories are also confirmed when we simulate the full eq.~\eqref{eq:glv} for a large enough number of species (see Supplementary Methods). 

We highlight that such divergences are not because the DMFT does not hold for a specific set of parameters (in fact simulating both eq.~\eqref{eq:glv} and eq.~(\ref{eq:dmft}) give rise to the same finite time divergence - see Supplementary Methods). Rather, the divergence at finite times and its dependence on initial conditions of the AGLV dynamics with $\tau=0$ are due to unbounded growth of the function $J(x)=x$ in the GLV model and non-linearity of the Fokker-Planck equation \cite{frank2005nonlinear}. However, when the Monod functional response is introduced through a bounded $J(x)$), we prevent uncontrolled population growth, eliminating any divergences and dependencies on the initial conditions previously observed. Therefore, the phase diagram displays only the stable stationary state for any values of $\mu$ and $\sigma$. The stationary solution, eq.\eqref{eq:PstazUCNA}, does not depend on the specific form of $J(x)$. The specific choice of $J(x)$ enters only through the self consistencies.

\begin{figure}[t]
\includegraphics[width=0.6\columnwidth]{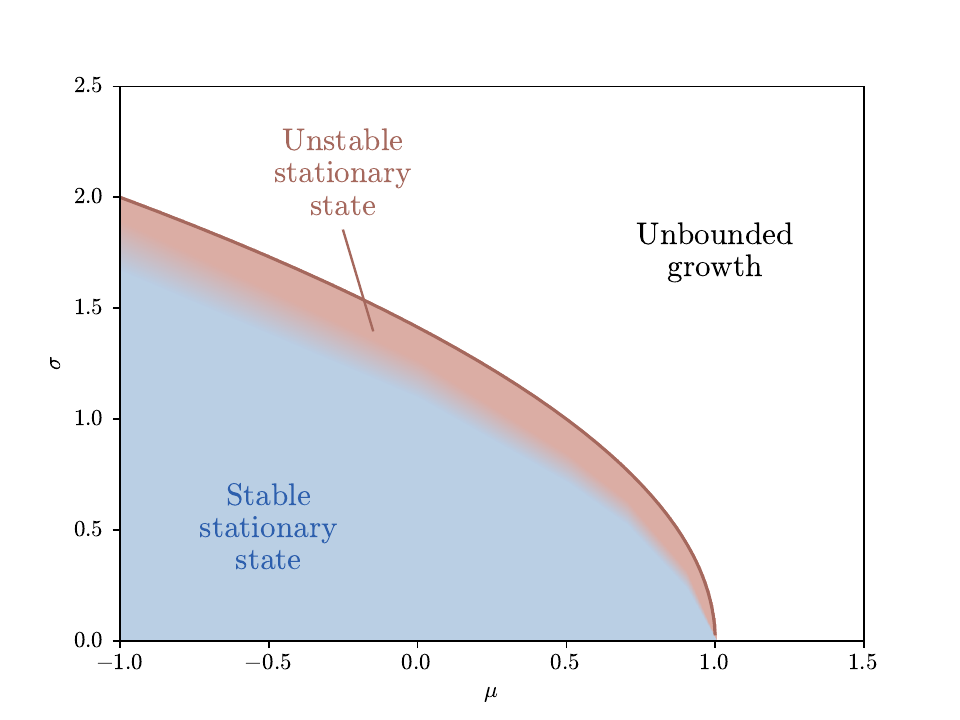}
\caption{\label{fig4} PhaseDiagram for the annealed white noise case and $J(x)=x$. We show the AGLV phase diagram as a function of the mean ($\mu$) and standard deviation ($\sigma$) of the species' interaction strengths. In the unbounded growth region, as predicted analytically, the species abundance dynamics diverge for finite times. In the red-purple region, although the stationary solution of the DMFT equation exists, it is not reachable by the dynamics and we also have a singularity for finite times. In the light-blue region, for initial conditions taken from the stationary distribution eq.~\eqref{eq:PstazAnnS}, a stationary state exists and it corresponds to the Gamma distribution as given by eq.~\eqref{eq:PstazAnnS}. The gradient between the blue and brown regions indicates a region of numerical uncertainty. Divergences and numerical uncertainty are eliminated when $J(x)=x/(1+cx)$ is considered in the dynamics, and the corresponding phase diagram only displays a stable stationary state.}
\end{figure}

In this study, we have undertaken an investigation into the GLV equations with annealed disorder, incorporating finite correlation time and simple functional responses. We have determined the corresponding dynamical mean-field equations for a large number of species, which do not depend on the specific form of $J(x)$. The inclusion of temporal stochastic fluctuations in the strengths of species interactions has resulted in a remarkably diverse range of phenomena and ecologically significant outcomes.
Firstly, the introduction of annealed disorder in the GLV equations, for any finite correlation time, has exerted a substantial positive influence on the biodiversity of the system. Specifically, when the dynamics of the system converge to the stationary distribution, we observe the quasi-cycles of species populations dynamics, where species abundances alternate between high and low values, favoring the coexistence of all species (if we do not artificially introduce any minimal threshold under which we consider the species extinct). This is, in fact, a similar outcome to what QGLV models found in the chaotic phase \cite{pearce2020stabilization,dePirey2023} when introducing an immigration rate $\lambda$. 
Second, in the white noise limit, the DMFT leads to the stochastic logistic model, a phenomenological model that proved to be consistent with several macro-ecological laws in microbial ecosystems \cite{grilli2020macroecological,descheemaeker2020stochastic}.
In particular, the analytical species abundance distribution derived from the DMFT follows the Gamma distribution, a widely utilized probability distribution in macroecology \cite{azaele2006dynamical,azaele2016statistical}. Again, similar truncated fat-tailed distribution has been recently shown in the chaotic phase \cite{dePirey2023} and in the strongly interacting limit \cite{hu2022emergent,mallmin2023chaotic}  of the QGLV with immigration. 
We have successfully obtained the phase diagram for the case of annealed white noise (AWN), and numerical simulations for the case $J(x)=x$ have revealed the potential for unbounded growth when the initial conditions possess large values, despite the existence of an analytically stationary solution. In other words, due to the non-linear nature of the corresponding Fokker-Planck equation and known pathologies in the GLV model (also observed in the quenched case), the dynamics may not converge to the stationary solution, leading to divergent trajectories. 
Eventually, we have presented a refinement of the model, through the inclusion of a simple functional response. We have shown that it not only maintains the core phenomenology described above, but also rectifies any non-physical divergences. Thus, it enhances the model's realism and applicability without sacrificing its fundamental characteristics and predictive capabilities.
This work opens various other avenues of research, including the integration of  quenched as well as annealed disorder and the correlations between pairs of interacting species \cite{bunin2017ecological,galla2018dynamically} or more complex hierarchical correlation structures \cite{poley2023generalized}.
More generally, the methodology presented here can be exploited to study the effect of annealed disorder also in other ecological dynamics. Moreover, to further improve the AGLV model, it would be valuable to explore sparse interaction networks instead of the fully connected ones examined here and in previous works \cite{barbier2018generic, biroli2018marginally, galla2018dynamically, pearce2020stabilization, bunin2017ecological, altieri2021properties} for the quenched version. However, unlocking the dynamics of this intriguing scenario requires a more comprehensive generalization of the DMFT approach outlined in this study \cite{azaele2023}.
The exploration of such directions holds significant promise for advancing the modeling of large-scale ecosystem dynamics, understanding emergent macro-ecological patterns observed in empirical data, and investigating the influence of environmental fluctuations on species coexistence.


\begin{acknowledgments}
We wish to acknowledge Jacopo Grilli and Davide Bernardi for critical reading of the manuscript and useful discussions. F.F. thanks Matteo Guardiani and the Information Field Theory group at the Max-Planck Institute for Astrophysics for their hospitality and helpful comments. S.S. acknowledges Iniziativa PNC0000002-DARE - Digital Lifelong Prevention. S.A., F.F., and A.M. also acknowledge the support by the Italian Ministry of University and Research (project funded by the European  Union - Next Generation EU: “PNRR Missione 4 Componente 2, “Dalla ricerca all’impresa”, Investimento 1.4, Progetto  CN00000033”).
\end{acknowledgments}

\bibliography{glv-bib}

\end{document}